\documentstyle[prl,aps]{revtex}
\begin{document}
\draft
\twocolumn
\newcommand{\emi}{{\rm em}}
\newcommand{\abs}{{\rm abs}}
\newcommand{\tot}{{\rm tot}}

\topmargin 0in
\title{Delay-dependent amplification of a probe pulse via stimulated Rayleigh scattering}

\author{M. V. Fedorov}
\address{General Physics Institute, Russian Academy of Sciences, 
38 Vavilov st., 119991, Moscow, Russia}
\author{S. V. Popruzhenko}
\address{Moscow Engineering Physics Institute, Kashirskoe Shosse 31,
115409, Moscow, Russia}
\author{D. F. Zaretsky}
\address{Russian Research Center ``Kurchatov Institute'', 123182, Moscow, 
Russia}
\author{W. Becker\thanks{also at Center for Advanced Studies, Dept. of Physics 
and Astronomy, Univ. of New Mexico, Albuquerque, NM 87131.}}
\address{Max-Born-Institut, Max-Born-Str. 2a, 12489 Berlin, Germany}

\date{\today}

\maketitle
\begin{abstract}

Stimulated Rayleigh scattering of pump and probe light pulses of close
carrier frequencies is considered. A nonzero time delay between
the two pulses is shown to give rise to amplification of the delayed 
(probe) pulse 
accompanied by attenuation of the pump, both on resonance and 
off resonance. In either case, phase-matching effects are shown to 
provide a sufficiently large gain, which can exceed significantly direct 
one-photon-absorption losses. 
\end{abstract}


\maketitle

Rayleigh scattering in continuous media is attributed to 
scattering off ``nonpropagating modulations of material observables'' 
\cite{Courtois}. The elementary quantum-mechanical process underlying  
Rayleigh scattering is the two-photon emission-absorption transition of an 
atom such that the initial and final atomic states coincide  
\cite{Ray_def}. In the approximation where both atomic recoil and motion of 
the atoms are ignored,  the frequencies of the emitted and absorbed photons 
are equal though the directions of their wave vectors can be different. The 
concept of stimulated Rayleigh scattering arises when one considers the same 
atomic transitions but in the presence of two specified waves, pump and probe, 
such that stimulated emission of either pump or probe photons prevails over 
spontaneous emission.  

In a traditional formulation of the problem of stimulated Rayleigh scattering 
\cite{Courtois}, both pump and probe wave are assumed to be monochromatic plane waves. 
In this case, for an ensemble of stationary recoilless atoms, there is no net Rayleigh scattering: the probability of  the direct process (absorption
of a pump photon and emission of a probe photon) is exactly equal to that of   the inverse process (emission of a pump photon and
absorption of a probe photon). Hence, there is no net conversion of the two species of photons into each other. Here, we will consider a different situation: we 
will assume that pump and probe fields are finite pulses,  whose centers are separated by the variable delay $\Delta t$. It will be shown that in this case a photon of the pump wave can be transformed 
into a photon of the (delayed) probe wave. The efficiency of this process can be very high, owing to phase matching effects of coherent 
stimulated Rayleigh scattering off an ensemble of atoms.

In the case we consider, the phases of pump and probe waves are completely 
uncorrelated. Owing to this assumption, only the 
{\em probabilities} of the  direct and inverse 
processes described above are meaningful quantities, rather than a  coherent sum of
transition {\em amplitudes}. 
The complete lack of coherence between the pump and the probe wave distinguishes this case 
from  pulse propagation in a resonant medium \cite{Allen,Crisp} (otherwise,  the pump and the probe could be combined into one pulse whose propagation one could study). Also, evidently, stimulated Rayleigh scattering is conceptually different from 
bichromatic pulse propagation in three-level media, where the elementary 
underlying processes are $\Lambda$ or Raman transitions between different 
initial and final atomic states \cite{Hioe,Shore}. 

It should also be mentioned  that in our case  the pump-to-probe
transformation of photons or, in other words, the amplification of the probe wave is not connected 
with any kind of population inversion: initial- and final-state 
populations are 
equal because initial and final states are identical. In this sense,  
stimulated Rayleigh scattering is reminiscent of the well-known inversionless lasers 
\cite{Kochar}, even though the physics of amplification in our scheme and in 
inversionless lasers are completely different. 
 
A  closely related problem was investigated in a recent
work on amplification of high-order harmonics of a strong laser field
\cite{HHG}. However, the latter is
a rather complicated process, whose theoretical description is based
on various assumptions and approximations, which shadow the physics of
the phenomenon. On the other hand, the problem formulated above is
so fundamental that it deserves description of its simplest manifestation,
which occurs in Rayleigh 
scattering to be considered in this letter. 

So, let the electric-field strength of the pump and the probe pulse  be given by

\begin{eqnarray}
\varepsilon(t)&=&\varepsilon_0(t)\cos(\omega_0 t-{\bf k} \cdot {\bf r}),\nonumber\\
 \tilde{\varepsilon}(t)&=&\tilde{\varepsilon}_0(t)\cos(\tilde{\omega}_0 t
 - \tilde{\bf k} \cdot {\bf r}), \label{fields}
\end{eqnarray}
respectively, where 
$|{\bf k}|=\omega_0/c$ and
$|\tilde{\bf k}|=\tilde{\omega}_0/c$. The fields of both pulses are
assumed to be linearly polarized along the same direction. 

By assuming that an atom, located at ${\bf r}={\bf R}_j$, is
initially in its ground state and the fields $\varepsilon$ and
$\tilde{\varepsilon}$ are weak enough, we calculate the
second-order perturbation-theory  amplitude of the
transitions involved in stimulated Rayleigh scattering and present
it in the form
\begin{equation}
 \label{total-amplitude}
 A^j=A_{\emi}^j+A_{\abs}^j,
\end{equation}
where $A_{\emi}^j$ and $A_{\abs}^j$ are, respectively, the
transition amplitudes involving
stimulated emission and absorption of a probe-wave photon
\cite{HHG,ZAR}
\begin{equation}
 \label{em/abs-j}
 A_{\left\{\emi \atop{\abs}\right\}}^{j}=
 A_{\left\{\emi \atop{\abs}\right\}}\,
 \exp{[\,\pm i({\bf k}-{\tilde{\bf k}})\cdot {\bf R}_j\,]}
\end{equation}
and 
\begin{eqnarray}
A_{\left\{\emi \atop{\abs}\right\}}=&-&\sum_n |d_{0n}|^2
 \int_{-\infty}^{\infty}dt\int_{-\infty}^t dt^{\prime}
 e^{i(E_0-E_n)(t-t')}\nonumber\\
  &\times&\Bigl(\varepsilon_0(t)\tilde{\varepsilon}_0(t^{\prime})\exp[\mp i(\omega_0t-\tilde
 {\omega}_0t^{\prime})]\nonumber\\
 &+&\tilde{\varepsilon}_0(t)\varepsilon_0(t^{\prime})
 \exp[\pm i(\tilde{\omega}_0t-\omega_0t^{\prime})]\Bigr).
 \label{em/abs}
\end{eqnarray}
Atomic units are used throughout the paper if not indicated otherwise. In Eq.~(\ref{em/abs}), $d$ is the
projection of the atomic dipole moment upon the direction of
polarization, $d_{0n}$ is its matrix elements between the ground state 
$|0\rangle$ and an intermediate atomic state $|n\rangle$, and
the sum over $n$ includes integration over the atomic
continuum $E$. 

Identification of the two terms in the sum
(\ref{total-amplitude}) as  the amplitudes of stimulated emission and 
absorption is related to the  sign in front of the carrier frequency $\tilde\omega_0$ of the probe 
in the  exponent on the right-hand side of Eq.~(\ref{em/abs}). 
This separation of emission and absorption is illustrated by the four 
diagrams of Fig.~1. 
The first two  diagrams 
of Fig.~1 [(a) and (b)] correspond to emission of a probe-wave photon with frequency 
$\tilde{\omega}$ (the arrows of the lines pointing down), whereas the 
other two  [(c) and (d)] correspond to absorption (the arrows 
 pointing up). 
The two diagrams in each line of Fig. 1 correspond to the two terms in 
big parentheses on the right-hand side of Eq. (\ref{em/abs}). 
Only the near-resonant terms (b) and (c) make a significant contribution.
These two terms are proportional to
$\tilde{\varepsilon}_0(t)\varepsilon_0(t')$  and
$\varepsilon_0(t)\tilde{\varepsilon}_0(t')$, respectively. 
Since in Eq.~(\ref{em/abs}) we always have $t>t'$, already this very general 
expression shows that stimulated emission can prevail over absorption provided 
the probe pulse is retarded with respect to the pump pulse, because in
this case $\tilde{\varepsilon}_0(t)\varepsilon_0(t')
>\varepsilon_0(t)\tilde{\varepsilon}_0(t')$, if pulse durations are
short enough.

By expanding $\varepsilon(t)$ and $\tilde{\varepsilon}(t)$ 
in terms of Fourier integrals
\begin{equation}
 \label{Fourier}
 \left\{\varepsilon(t)\atop{\tilde{\varepsilon}(t)}\right\}
 =\int_{-\infty}^{\infty}d\omega\,\
 \exp{(i\omega t)}\left\{\varepsilon_{\omega}\atop{\tilde{\varepsilon}_{\omega}}\right\}
,
\end{equation}
we can reduce Eq. (\ref{em/abs}) to the much simpler form
\begin{equation}
 \label{em-abs-pol}
 A_{\left\{{\emi \atop {\abs}}\right\}}=2\pi
 i\,\int_0^{\infty}d\omega\,\alpha(\omega)
 {\left\{{\varepsilon_{\omega}^*\tilde{\varepsilon}_{\omega}
 \atop {\varepsilon_{\omega}\tilde{\varepsilon}_{\omega}^*}}\right\}},
\end{equation}
where $\alpha(\omega)$ is the complex atomic polarizability
$$
 \alpha(\omega)\equiv\alpha_1(\omega)+ i\,\alpha_2(\omega)=\sum_n\,|d_{0n}|^2
$$
\begin{equation}
 \label{polarizability}
 \times\,\left(\left.\frac{1}{E_n-E_0-\omega-i\delta}
 +\frac{1}{E_n-E_0+\omega-i\delta}\right)\right|_{\delta\rightarrow
 0}.
\end{equation}

For a gas of atoms two-photon absorption-emission transitions in 
different atoms correspond to the same initial and final state but different 
intermediate states of the total multiatomic system. Summation over 
intermediate states includes summation of the probability amplitudes 
$A^j_{\emi / \abs}$ of the various atoms. The total probability of 
emitting a probe photon $\tilde{\omega}$ is given by the difference of the probabilities 
of its stimulated emission and absorption summed over all atoms,
\begin{equation}
 \label{w-T-tot}
 w_T^{\tot}=\Big|\sum_jA_{\emi}^j\Big|^2-\Big|\sum_jA_{\abs}^j\Big|^2 \equiv w_T F,
\end{equation}
where $F$ is the phase-matching factor \cite{Agostini,ZAR,HHG}
\begin{equation}
 \label{F-factor}
 F=\Big|\sum_j
 \exp{[\,i({\bf k}-{\tilde{\bf k}})\cdot {\bf R}_j\,]}\Big|^2
\end{equation}
and $w_T$ the single-atom total emission probability, $w_T =|A_{\emi}|^2
 -|A_{\abs}|^2$.

If $w_T\;{\rm and}\;w_T^{\tot}$ are positive, the gain $G$ of the
probe wave in an atomic gas is determined as the ratio of the average total 
energy $\hbar \omega_0 w_T^{\tot}$ gained by the probe wave over its incident energy $V\tilde{\varepsilon}_0^2/8\pi$, 

\begin{equation}
 \label{gain}
 G=\frac{8\pi\,
 \hbar\tilde{\omega}_0}{V\,\tilde{\varepsilon}_0^2}\,w_T^{\tot}=
 \frac{8\pi\,
 \hbar\tilde{\omega}_0}{\tilde{\varepsilon}_0^2} n_a \frac{F}{N_a}\,w_T,
\end{equation}
where $V$ is the interaction volume, $N_a$ the total number of atoms  and $n_a=N_a/V$ their density. 

The phase-matching factor (\ref{F-factor}) arises from 
coherent (in-phase) emission of photons by different atoms, and it
can be rather large. 
For ${\bf k} = \tilde{\bf k}$, it assumes its maximal value  
$F_{\max} =N_a^2$. However, in order that the pump and the probe mode can be experimentally distinguished, we will consider ${\bf k} \approx \tilde{\bf k}$. If the length of the interaction
region in the direction of $\,{\bf k}-\tilde{\bf k}$ is 
$\leq 1/|{\bf k}-\tilde{\bf k}|$, we still have 
$F\approx F_{\max}$. In that case, the
coherent gain (\ref{gain}) differs from the
incoherent one by the factor $F/N_a \gg 1$.

Let us now specialize the pulse envelopes $\varepsilon_0(t)$ and
$\tilde{\varepsilon}_0(t)$ to  the Gaussians
\begin{equation}
 \label{Gauss}
 \varepsilon_0(t)=\varepsilon_0\,\exp{\left(-\frac{t^2}{2\,\tau^2}\right)},
 \,\tilde{\varepsilon}_0(t)
 =\tilde{\varepsilon}_0\,\exp{\left(-\frac{(t-\Delta t)^2}
 {2\,\tilde{\tau}^2}\right)},
\end{equation}
where $\Delta t$ is the delay of the probe pulse.  For simplicity, we assume that the pump and the probe pulse have identical  durations and carrier frequencies, 
$\tilde{\omega}=\omega\equiv\omega_0$ and $\tilde{\tau}=\tau$. Then, 
 the probability amplitudes (\ref{em-abs-pol})
become 

\begin{eqnarray}
 A_{\left\{\emi \atop{\abs}\right\}}=i\,
 \frac{\varepsilon_0\,\tilde{\varepsilon}_0\,\tau^2}{4}
 \int_0^{\infty}d\omega\,\alpha(\omega)\nonumber\\
 \times\,\exp{\left[-(\omega-\omega_0)^2\tau^2
 \mp i\,(\omega-\omega_0)\Delta t\right]}.\label{ampl-polar-Gauss}
\end{eqnarray}

First, we will assume that the carrier frequencies of both pulses are 
resonant with 
some bound-bound atomic transition, $\tilde{\omega}_0=\omega_0=E_1-E_0$. 
In this case, the polarizability
(\ref{polarizability}) can be approximated by its resonant part

\begin{equation}
 \label{polariz-res}
 \alpha(\omega)\approx -\frac{|d_{10}|^2}
 {(\omega-\omega_0)+\frac{i}{2}\Gamma},
\end{equation}
where $\Gamma$ is the width of the level $E_1$, which we put in by hand (a calculation to all orders in the pump would furnish it automatically). Under these
conditions, directly from Eqs. (\ref{ampl-polar-Gauss}) and
(\ref{polariz-res}) we get

\begin{equation}
 \label{wT-res-final}
 w_T=(\tau\Omega_R)^2\,(\tau\tilde{\Omega}_R)^2\,J,
\end{equation}
where $\Omega_R=\frac{1}{2}\varepsilon_0 |d_{01}|$ and
$\tilde{\Omega}_R=\frac{1}{2}\tilde{\varepsilon}_0 |d_{01}|$ are
the pump- and probe-wave Rabi frequencies, and $J$ is a
dimensionless function of the dimensionless variables
$t_0=\Delta t/\tau$ and $\gamma=\Gamma\tau$,
\begin{equation}
 \label{J-exact-res}
 J=-\gamma\frac{d}{dt_0}\left|\int_{-\infty}^{\infty}dx
 \frac{\exp(-x^2+it_0x)}{x^2+\gamma^2/4}\right|^2.
\end{equation}

For $\gamma\gg 1$,  
Eqs.~(\ref{wT-res-final})--(\ref{J-exact-res}) yield

\begin{equation}
 \label{large width}
 w_T=\frac{16\,\pi\,\Omega_R^2\,\tilde{\Omega}_R^2}{\Gamma^3}\,\Delta
 t\,\exp\left\{-\frac{1}{2}\left(\frac{\Delta t}{\tau}\right)^2\right\}.
\end{equation}
Increasing the width $\Gamma$ results in a rather quick
decrease of the two-photon Rayleigh-scattering probability
$w_T$. This restricts (for $\Gamma\tau >1$) the
region of time delays $\Delta t$ where $w_T$ is not small to
the range $|\Delta t|\lesssim 1/\tau$ (cf. Fig. 2). 

The antisymmetric shape (with respect to $\Delta t$) of the gain (\ref{large width}) is reminiscent of the free-electron laser (FEL), whose gain is antisymmetric with respect to the detuning \cite{madey}. There are other interesting relations between our scheme and the FEL that will be discussed elsewhere.

It should be noted that  the width
 $\Gamma$ can be significantly larger than a typical 
single-atom radiative width $\Gamma_r\sim 10^8 {\rm s}^{-1}$, owing to coherent   spontaneous forward emission by coherently excited atoms. 
Actually, in addition to the two-photon transition we here consider, the pump 
field also provides a real population of the resonant level $E_1$. 
Atoms excited 
in such a way can emit photons spontaneously. This is a sequential two-step 
absorption and emission process, which is  different from the coherent quantum-mechanical 
two-photon transition considered above. But, if the phases of the 
atomic excitation 
do not change significantly during the duration of the pulse, spontaneous 
emission is characterized by the same phase-matching factor (\ref{F-factor}) as stimulated emission. For the former, however, the wave vector 
$\tilde{\bf k}$ can have arbitrary direction. For this reason, the 
phase-matchimg factor for spontaneous emission appears to be given by 
$F_{\rm sp}=\int F({\bf n})d\Omega_{\bf n}$, where 
${\bf n}=\tilde{\bf k}/\tilde k$. In other words, laser excitation prepares a
coherent Dicke-type ensemble of excited atoms, whose spontaneous emission  can 
be strongly enhanced in comparison with the case of incoherently excited 
atoms \cite{Allen,Dicke}. For a laser focus with waist $d$ and  
length $L\sim d^2/\lambda$, where $\lambda=2\pi/k$ is the wavelength, a simple 
estimate based on replacing the summation in Eq.~(\ref{F-factor}) by 
an integration gives $F_{\rm sp}\sim (d/L)^2N_a^2$. The product 
$\Gamma_rF_{\rm sp}$ determines the total number of photons emitted per unit time by 
the ensemble of atoms in the focal volume. Divided by $N_a$ it gives 
the rate of
transitions per single atom or a phase-matching-modified radiative width
of the level $E_1$: $\Gamma=\Gamma_r F_{\rm sp}/N_a\sim \Gamma_r (d/L)^2N_a$. 
For example, for $\lambda\sim 10^{-4}$ cm, $d\sim 10^{-3}$ cm, 
$L\sim 10^{-2}$ cm, and $n_a\sim 10^{16}$ cm$^{-3}$ we get $N_a\sim 10^8$, 
$F_{\rm sp}\sim 10^{14}$ and $\Gamma\sim 10^{14}$ s$^{-1}$. Therefore, 
even for 
100-fs pump pulses the parameter $\Gamma\tau\simeq 10\gg 1$,
and the approximation (\ref{large width}) is relevant.
For these parameters and for $\Delta t\sim\tau\sim 10^{-13}{\rm s}$, 
$\tilde{\Omega}_R\tau\sim \Omega_R\tau\sim 10^{-2}$,   
the gain (\ref{gain}) of the probe wave is of the order of one, $G\sim 1$. 
Finally, as the direct one-photon absorption of probe photons does not 
experience any phase-matching enhancement, it can be checked easily to be much 
less efficient than the here described two-photon stimulated Rayleigh 
scattering process.

Next, we will consider the case when the pump- and probe-wave
frequencies $\omega$ and $\tilde{\omega}$ exceed the
ionization energy $E_b$.
For direct numerical calculations with the help of Eqs.
(\ref{w-T-tot}) and (\ref{ampl-polar-Gauss}) we use the
$\delta$-potential model, for which the atomic polarizability
is  given by \cite{Demkov} 
$\alpha(\omega)=a(x)/E_b^2$, where $x=\omega/E_b$ and
\begin{equation}
 \label{a(x)}
 a(x)=-\frac{1}{x^2}+\frac{8}{3x^4}-
 \frac{4}{3}\frac{(x+1)^{3/2}-i\,(x-1)^{3/2}}{x^4}.
\end{equation}

In the case $\omega_0-E_b \gg 1/\tau$ the integral over $\omega$ in 
Eq.~(\ref{ampl-polar-Gauss}) can be calculated analytically to give

\begin{eqnarray}
w_T &=&\frac{\pi\,\varepsilon_0^2\,\tilde{\varepsilon}_0^2}{8}
\Delta t \,\exp{\left(-\frac{\Delta t^2}{2\tau^2}\right)} \nonumber\\
 &\times& \Bigl(\alpha_1(\omega_0)\alpha_2^{\prime}(\omega_0)
 -\alpha_2(\omega_2)\alpha_1^{\prime}(\omega_0)\Bigr). \label{w-cont}
\end{eqnarray}

The dependence of $w_T$ on the delay  $\Delta
t$ is similar to that occurring in the resonance case when the 
width of the resonant level is large (cf. Fig. 2): $w_T(\Delta t)$ is an odd
function and is localized in the range $|\Delta t|\lesssim 1/\tau$.

The dependence of $w_T$ [Eq.~(\ref{w-cont})] on $\omega_0$ is shown in
Fig. 3, and this result shows that the effect is maximal
approximately at $\omega_0\approx 1.2\,E_b$.

In the case of nonresonant transitions, the gain (\ref{gain}) can  
easily become as large as in the resonant case. 
For the same values of $\lambda$, $d$, and $L$ as in the previous estimates,
the value $G\sim 1$ is reached at  $n_a\sim 10^{17}$ cm$^{-3}$.
Again, as the direct losses due to ionization do not experience phase-matching 
enhancement, they are insignificant on the scale of the energy transfer 
that is 
achievable in  two-photon stimulated Rayleigh scattering.

In conclusion, stimulated Rayleigh scattering, both resonant with an atomic bound-bound transition or via the continuum, may furnish substantial gain in a pump-probe experiment where the probe pulse is delayed with respect to the pump.
This scenario appears to have escaped attention thus far.

MVF enjoyed discussions with W. P. Schleich. This work was partially supported  by the Russian Fund of Basic
Research (grants No 99-02-18034, 99-02-04021 and 00-02-17078), 
INTAS (grants No
99-450 and 99-01495), CRDF (grant No RP1-2259), and the Alexander von Humboldt Foundation.

Fig. 1. Diagrams of two-photon transitions involved in stimulated Rayleigh 
scattering. The ground state with energy $E_0$ (solid) and a near-resonant excited state with energy $E_n$ (dashed) are indicated by horizontal lines. The latter may as well represent the continuum threshold.
\vskip 3cm

Fig. 2. The function $J(\Delta t/\tau)$ [Eq.~(\ref{J-exact-res})], which is proportional to the 
total probability (\ref{wT-res-final}) of resonant stimulated Rayleigh scattering  for $\Gamma\tau=10$.
\vskip 3cm

Fig. 3. The function ${\rm Re}\,a(x){\rm Im}\,a'(x)-{\rm Im}\,a(x){\rm Re}\,a'(x)$, which  characterizes  
the dependence of $w_T(\omega)$ [Eq.~(\ref{w-cont}].

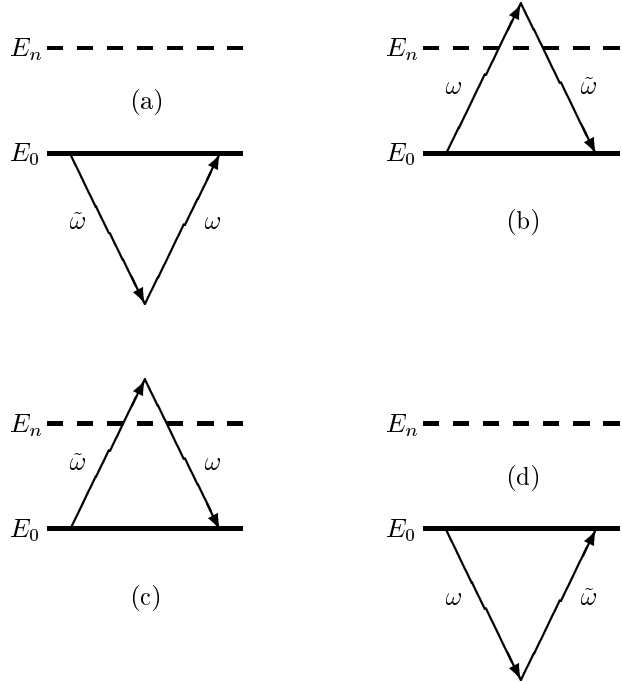
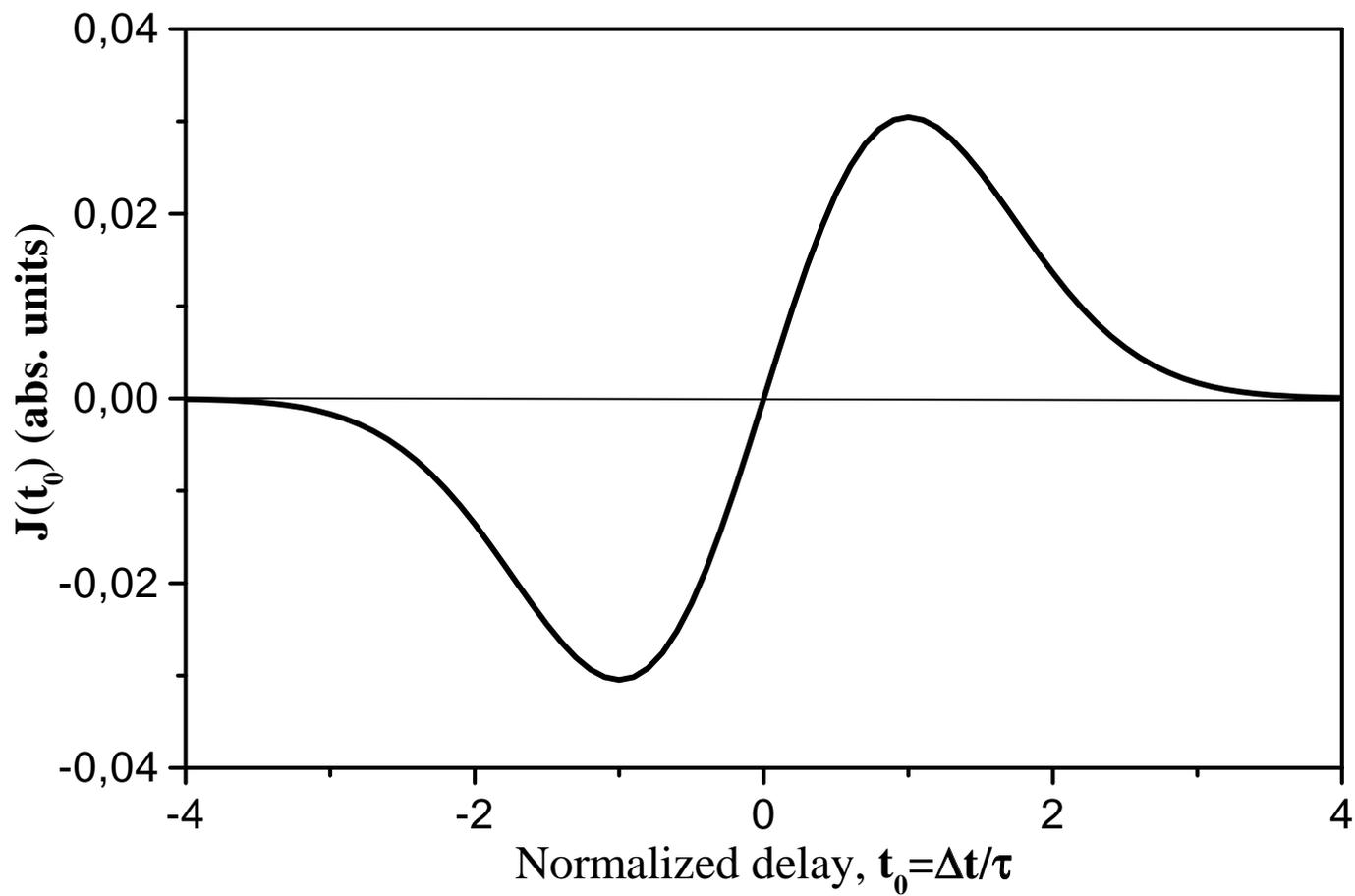
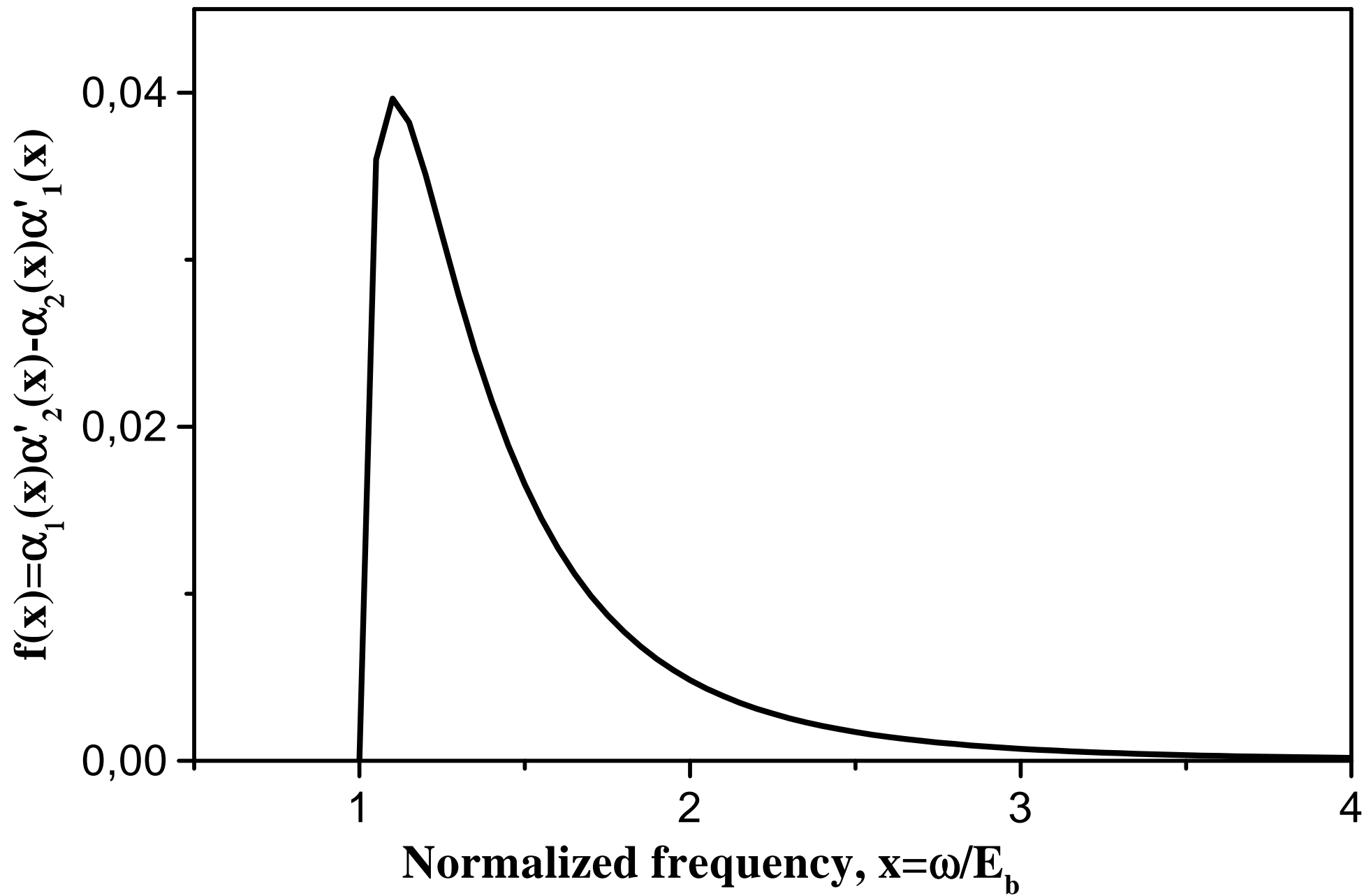
\end{document}